\documentclass[aps,jmp,reprint,prl,showpacs]{revtex4-1}
\usepackage{graphicx}
\usepackage{amsmath}
\usepackage{amssymb}
\usepackage{dcolumn}
\usepackage{multirow}
\usepackage{hhline}
\usepackage{blindtext}
\usepackage[acronym]{glossaries}
\usepackage[utf8]{inputenc}
\usepackage{braket}
\usepackage{color}
\newacronym{2deg}{2DEG}{two-dimensional electron gas}
\newacronym{qpc}{QPC}{quantum point contact}
\newacronym{qd}{QD}{quantum dot}
\newacronym{awg}{AWG}{arbitrary waveform generator}
\newacronym{dos}{DOS}{density of states}

\newcommand{\win}{\ensuremath{W_{\textnormal{0,in}}}}
\newcommand{\wout}{\ensuremath{W_{\textnormal{0,out}}}}

\newcommand{\gout}{\ensuremath{\Gamma_\textnormal{out}}}
\newcommand{\gin}{\ensuremath{\Gamma_\textnormal{in}}}

\newcommand{\ees}{\ensuremath{E_\textnormal{es}}}

\begin{document}
\preprint{ETH Zurich}

\title{Measuring the degeneracy of discrete energy levels using a GaAs/AlGaAs quantum dot} 

\author{A. Hofmann}
\email[]{andrea.hofmann@phys.ethz.ch}
\author{V. F. Maisi}
\author{C. Gold}
\author{T. Kr\"ahenmann}
\author{C. R\"ossler}
\author{J. Basset}
\author{P. M\"arki}
\author{C. Reichl}
\author{W. Wegscheider}
\author{K. Ensslin}
\author{T. Ihn}
\affiliation{Laboratory for Solid State Physics, ETH Zurich}

\date{\today}

\begin{abstract}
We demonstrate an experimental method for measuring quantum state 
degeneracies in bound state energy spectra.
The technique is based on the general principle of detailed balance, 
and the ability to perform precise and efficient measurements of 
energy-dependent tunnelling-in and -out rates from a reservoir. The 
method is realized using a GaAs/AlGaAs quantum dot allowing for the 
detection of time-resolved single-electron tunnelling with a precision 
enhanced by a feedback-control. It is thoroughly tested by tuning orbital 
and spin-degeneracies with electric and magnetic fields.
The technique also lends itself for studying the connection between the
ground state degeneracy and the lifetime of the excited states.
\end{abstract}

\maketitle 

Degeneracies play an important role in quantum statistics
\cite{huang_statistical_1988}. They often arise from symmetries of the underlying system
\cite{noether_invariant_1971,el-batanouny_symmetry_2008} 
and govern the theoretical description of macroscopic quantum 
phenomena such as superconductivity 
\cite{buckel_superconductivity:_2004} 
and the quantum Hall effect 
\cite{prange_quantum_1990}, 
but also play an important role for atomic spectra 
\cite{landau_course_1981}. 
Theoretical concepts of topological protection are based on ground-state degeneracies 
\cite{freedman_topological_2003}, 
and modern schemes to control qubits make use of tunable degeneracies 
\cite{nielsen_quantum_2010}. While the concept is omnipresent in quantum theory, measuring the 
degeneracy of an energy level in a quantum system seems to be less 
developed. A familiar way to experimentally demonstrate the existence 
of a degeneracy consists in breaking underlying symmetries, thereby lifting 
the degeneracy as in the Zeeman- 
\cite{tarucha_shell_1996,sasaki_spin_1998,lindemann_stability_2002,hanson_semiconductor_2004}, 
or the Stark-effects. Alternative techniques use selective excitations such as left- or 
right-circularly polarized light to distinguish degenerate excitations 
\cite{miah_photo-induced_2014}.

We demonstrate an experimental method of measuring the degeneracy 
of discrete energy levels alternative to the techniques 
mentioned above. The method is based on a general relation derived 
from detailed balance, and makes use of tunnelling spectroscopy and our ability 
to detect individual tunnelling events in real time 
\cite{schleser_time-resolved_2004}. We overcome previous accuracy limitations of this technique 
\cite{gustavsson_electron_2009} 
by implementing a feedback-control. A single few-electron quantum dot in 
GaAs serves as the system of choice to test our experimental method. In this 
system, ground and excited states are well studied
\cite{tarucha_shell_1996,kouwenhoven_excitation_1997,cooper_direct_2000,gould_correlations_2000,ciorga_addition_2000,potok_spin_2003,sasaki_electrical_2005, meunier_experimental_2007,amasha_electrical_2008,xiao_measurement_2010} 
and the presence of degeneracies is established from symmetry-breaking measurement techniques 
\cite{ashoori_textitn_1993,tarucha_shell_1996,kouwenhoven_excitation_1997,ciorga_addition_2000,folk_ground_2001,luscher_investigation_2001,luscher_signatures_2001,lindemann_stability_2002,potok_spin_2003,hellmuller_spin_2015}. Our method reliably traces these degeneracies with great accuracy. Furthermore, 
the system combined with our measurement method allows us to controllably alter 
the degeneracy of energy levels. The method of degeneracy detection is very 
general and can be directly transferred to other systems where states are 
accessible by tunnelling.

\begin{figure}
\includegraphics[width=\linewidth]{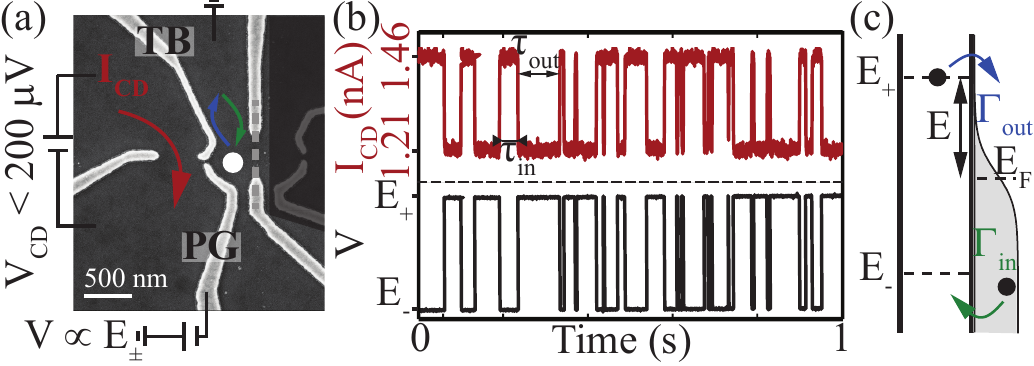}
\caption{Measurement setup. (a) the surface of the crystal
with Ti/Au top-gates in bright grey and the two-dimensional
electron gas underneath the dark area. An electron (white circle) 
tunnels back-and-forth between the quantum
dot and the reservoir (blue and green arrows).
The dotted line indicates a closed barrier and
biased gates are plotted in brighter color than grounded gates.
(b) the current through the nearby quantum point contact 
as a function of time (red) measures the occupation of the quantum dot 
[$(N-1) \approx$~1.46~nA, $N \approx$~1.21~nA].
The voltage, converted to energy, applied to the gate PG 
(black) indicates the energy $E_\pm$ of the 
state $\mu_N$, as indicated in the 
energy diagram of the quantum dot-reservoir
system in (c).}
\label{fig:Setup}
\end{figure}

Our samples are made from a GaAs/AlGaAs heterostructure hosting a two-dimensional 
electron gas 90\,nm below the surface. As shown in Fig.~\ref{fig:Setup}(a) we form a quantum 
dot by applying negative voltages to the metallic top-gate fingers thereby depleting 
the electron gas below. The quantum dot is coupled by tunnelling to an electron 
reservoir at a temperature of $T\approx 50$\,mK. The presented measurements are performed
on two different samples, in different cooldowns and cryostats.

At fixed gate voltages single 
electrons tunnel back and forth between the dot and the reservoir 
if the addition energy $\mu_N$ for adding the $N$th electron to the quantum dot 
is within the thermal energy window of approximately
$3.5kT\approx 15\,\mu$eV 
($k$ is the Boltzmann constant)
around the reservoir Fermi energy
\cite{ihn_semiconductor_2009}. The addition energy $\mu_N$ is the energy difference between the 
$N$-electron and the $(N-1)$-electron many-body ground state energies, 
$E_N$ and $E_{N-1}$, of the quantum dot. Each of these two energies
can be degenerate, meaning that a number of microstates exists sharing 
the same $N$ and $E_N$. The dot occupation changes through 
single-electron tunnelling only between $N$ and $N-1$, i.e. by 
only one electron, owing to the Coulomb blockade effect
characterized by a charging energy of $1\,\mathrm{meV}\gg kT$. 
In our experiment we investigate small electron numbers $N<9$.

We measure the quantum dot occupation with the current $I_\textrm{CD}$ through
a quantum point contact charge 
detector [see Fig.~\ref{fig:Setup}(a)] coupled capacitively to the dot
\cite{schleser_time-resolved_2004,vandersypen_real-time_2004}. 
Time-resolved traces of $I_\textrm{CD}$ [see Fig.~\ref{fig:Setup}(b)] provide
the statistically distributed waiting times $\tau_\mathrm{in}$ for single-electron tunnelling from 
the reservoir into the dot, and $\tau_\mathrm{out}$ for tunnelling from the dot 
into the reservoir. An exponential speed-up compared to 
previous measurement schemes
\cite{schleser_time-resolved_2004,hellmuller_spin_2015} 
allows us to measure these waiting times 
when $\mu_N$ is detuned from the Fermi energy $E_\mathrm{F}$ by much more than 
$kT$, where tunnelling in one direction is exponentially suppressed due to the 
lack of occupied (empty) states in the reservoir for tunnelling-in (out). 
To speed up the measurement in the fast direction,
we implement a feedback mechanism to switch the level cyclically 
between $E_\pm=E_\mathrm{F}\pm E$ [see Figs.~\ref{fig:Setup}(b),(c)]. 
The waiting time for an electron to tunnel out at $\mu_N=E_\mathrm{F}+E$ 
gives an instance of $\tau_\mathrm{out}(E)$.
After the detection of the tunnelling-out event, we quickly switch
the empty level to $\mu_N=E_\mathrm{F}-E$, where we wait for an electron to tunnel in,
thereby measuring an instance of $\tau_\mathrm{in}(-E)$. The occupied level is switched
back to $E_\mathrm{F}+E$ where the cycle restarts. An 
electronic feedback triggers the voltage switch between $E_\pm$ when the 
respective change in occupation has been detected. Tunnelling rates 
$\Gamma_\mathrm{out/in}(\pm E)=\langle \tau_\mathrm{out/in}(\pm E)\rangle^{-1}$ 
are obtained by averaging the waiting times over time traces of ten seconds.

\begin{figure}
\includegraphics[width=\linewidth]{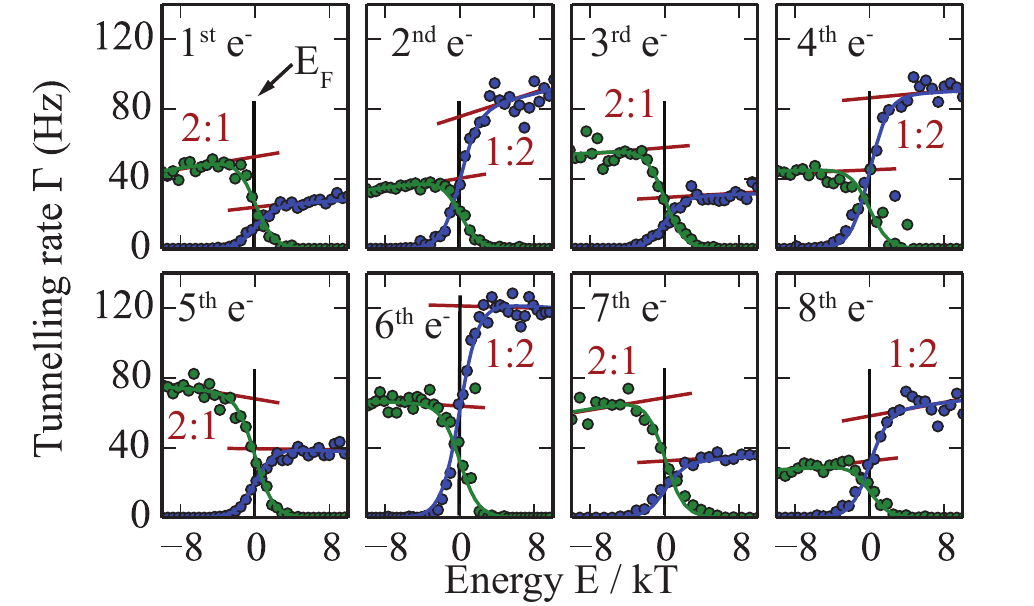}
\caption{Tunnelling rates of the first eight resonances around the
reservoir Fermi energy (zero energy reference). 
Green and blue color denote tunnelling-in and
tunnelling-out rate, respectively. The solid lines are fits
to Eq.~\eqref{eq:TunnelRates} and the red lines are guides to
the eye indicating the ratio $\win : \wout$ of the tunnel couplings 
to alternate between 2:1 and 1:2.}
\label{fig:EightElectrons}
\end{figure}

Measurements of $\Gamma_\mathrm{out/in}(E)$ are shown in Fig.~\ref{fig:EightElectrons} 
for tunnelling resonances corresponding to filling the first eight electrons into 
the quantum dot. We observe that tunnelling-in rates (green) are essentially 
constant for energies below resonance, usually with a weak linear energy-dependence superimposed
\cite{maclean_energy-dependent_2007}. 
Above resonance, these rates are exponentially suppressed according to the 
Fermi-distribution function. Conversely, tunnelling-out rates (blue) are 
essentially constant for energies above resonance (with a weak linear 
energy-dependence superimposed), and decrease exponentially for energies 
below. All measured curves in Fig.~\ref{fig:EightElectrons} agree with
\begin{align}
\label{eq:TunnelRates}
\begin{split}
\Gamma_\mathrm{in}(E) & = W_\mathrm{in}(E)f(E) \\
\Gamma_\mathrm{out}(E) & = W_\mathrm{out}(E)[1-f(E)],
\end{split}
\end{align}
with the Fermi-distribution function $f(E)$  and the energy-dependent tunnel 
couplings $W_\mathrm{in/out}(E)$.
Fitting the measured tunnelling-in and -out rates to Eqs.~\eqref{eq:TunnelRates}
with the the functions 
$W_\mathrm{in/out}(E)=W_\mathrm{0,in/out}(1+\alpha E)$
and the fitting parameters $W_\mathrm{0,in/out}$ and $\alpha$,
we determine the ratios of tunnelling rates on resonance $\win/\wout$. 
For the first eight resonances of the quantum dot, they alternate between 
integer ratios 2:1 and 1:2, as indicated in Fig.~\ref{fig:EightElectrons}.

This result may seem surprising in the light of the 
time-reversal symmetry of tunnelling. In this view, $\win=\wout$ is
expected, as the tunnel coupling of a given 
quantum dot state to the reservoir does not depend on the direction 
of tunnelling. However, this reasoning 
is incomplete because it neglects possible degeneracies of the initial and final 
quantum dot states involved in the tunnelling event
\cite{maclean_energy-dependent_2007, gustavsson_electron_2009}. 
One finds on the basis of detailed balance, that
\begin{equation}
\frac{W_{0,\mathrm{in}}}{W_{0,\mathrm{out}}} = \frac{p_N}{p_{N-1}} = \frac{m}{n},
\label{eq:DetailedBalance}
\end{equation}
where $m$ is the degeneracy of the $N$-electron and $n$ that of the 
$(N-1)$-electron energy level, and $p_N$ and $p_{N-1}$ are the 
time-averaged occupation probabilities of the two levels. 
Eq.~\eqref{eq:DetailedBalance} is the basic relation that allows 
us to determine the degeneracies of the different $E_N$ from the 
measurements in Fig.~\ref{fig:EightElectrons}.
It is likely that different orbitally degenerate states have different
tunnel couplings. Nevertheless, Eq.~\eqref{eq:DetailedBalance} is a ratio of 
integers, given only by the degeneracy of the initial and final states.
For weak energy dependence $\alpha$, this ratio can be read directly 
from the saturation values of the tunnelling rates at high and low energies.

For example, the resonance for filling the first electron in Fig.~\ref{fig:EightElectrons} 
is a transition between the the singly occupied dot ($N=1$) and the empty dot ($N-1=0$),
which has a non-degenerate energy $E_0=0$ leading to $n=1$. The measured ratio 
$\win : \wout = m:n =$2:1 indicates a two-fold degenerate ($m=2$) level $E_1$. 
It is well-known for this system
that indeed the $E_1$ state has a two-fold spin degeneracy 
\cite{fujisawa_allowed_2002}.
The resonance for filling the second electron is a transition between the 
two-fold degenerate $E_1$ state ($n=2$) and
the $E_2$ state. The measured ratio of $\win : \wout = m:n =$1:2 indicates 
a non-degenerate level $E_2$. This agrees with the well-known non-degenerate 
spin singlet ground state of the two-electron dot 
\cite{fujisawa_allowed_2002,hanson_semiconductor_2004,hanson_spins_2007,tarucha_shell_1996,ashoori_textitn_1993}, where two electrons with opposite spin occupy the lowest orbital single-particle state.
Along these lines we interpret the series of results in Fig.~\ref{fig:EightElectrons} 
as a sequence of alternating spin-up and spin-down filling into the quantum dot,
where energy levels $E_N$ are spin-degenerate for odd $N$, and non-degenerate 
for even $N$ up to $N=8$. This result 
fits well to the expectations in an asymmetric confinement potential
with orbitally non-degenerate single-particle states
\cite{ezaki_electronic_1997}.

\begin{figure}
\includegraphics[width=\linewidth]{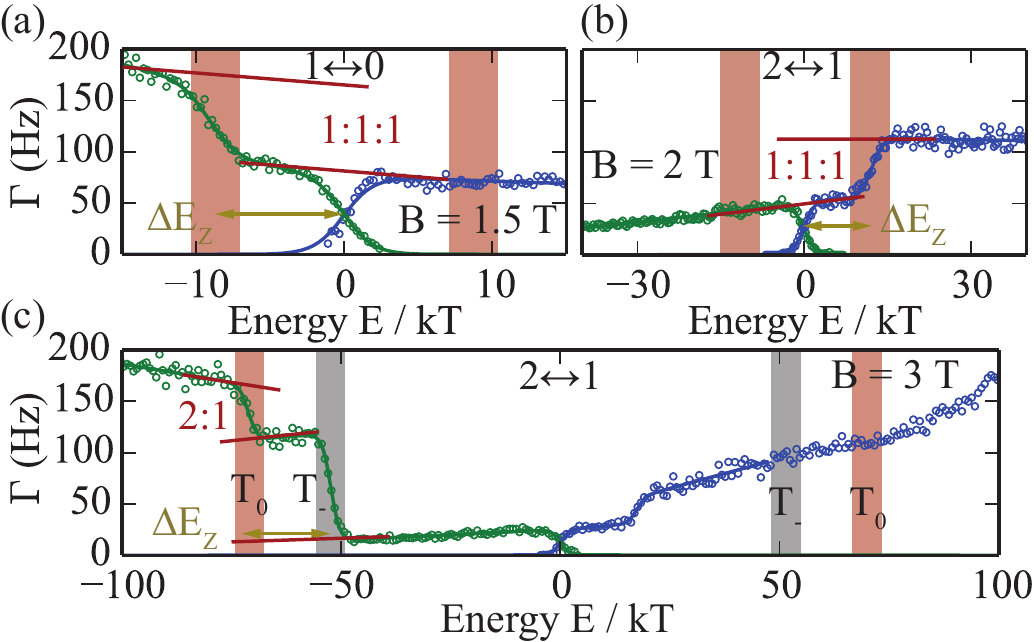}
\caption{Tunnelling rates at finite magnetic field
for the first (a) and second (b) resonance. The Zeeman energy
is $\Delta E_Z = g \mu_B B = |\ees|$ with the g-factor $|g|=0.44$
and the Bohr magneton $\mu_B$. The solid blue and green lines are 
fits to a modification of Eq.~\eqref{eq:TunnelRates} with adding energetically 
offset tunnelling rates. 
The red lines are guides to the eye and indicate the ratio 1:1:1 for
tunnelling into the ground-state, tunnelling-into the
Zeeman-split excited state and tunnelling-out of the ground
state. The ratios for the first and second electron are
$\wout :\win :\win^{\textrm{Z}}=$76.7:78.5:75.4
and $\wout :\win :\win^{\textrm{Z}}=$49.9:54.7:59.1,
respectively. (c) Excited state spectroscopy of the second electron.
The solid lines are fits to Eq.~\eqref{eq:TunnelRates} with
three (two) energetically offset tunnelling-in (-out) rates.
The red lines indicate the ratio 2:1 for
tunnelling into the triplet states $T_-$ and $T_0$,
$\win^{T_-}:\win^{T_0}=$105:54.}
\label{fig:BField}
\end{figure}

In order to further test the applicability of Eq.~\eqref{eq:DetailedBalance}
we use an in-plane magnetic field in order to lift the spin-degeneracy of the
one-electron ground state by the Zeeman effect.  The result is shown in 
Fig.~\ref{fig:BField}(a) for the tunnelling transitions between the empty 
and the one-electron dot, and in (b) for the transition between
the one- and two-electron ground state. The ratio of the ground state transition rates
now changed to $\win:\wout=$1:1 in both cases as predicted by
Eq.~\eqref{eq:DetailedBalance} for non-degenerate zero-, one- and two-electron
ground states.

\begin{figure}
\includegraphics[width=\linewidth]{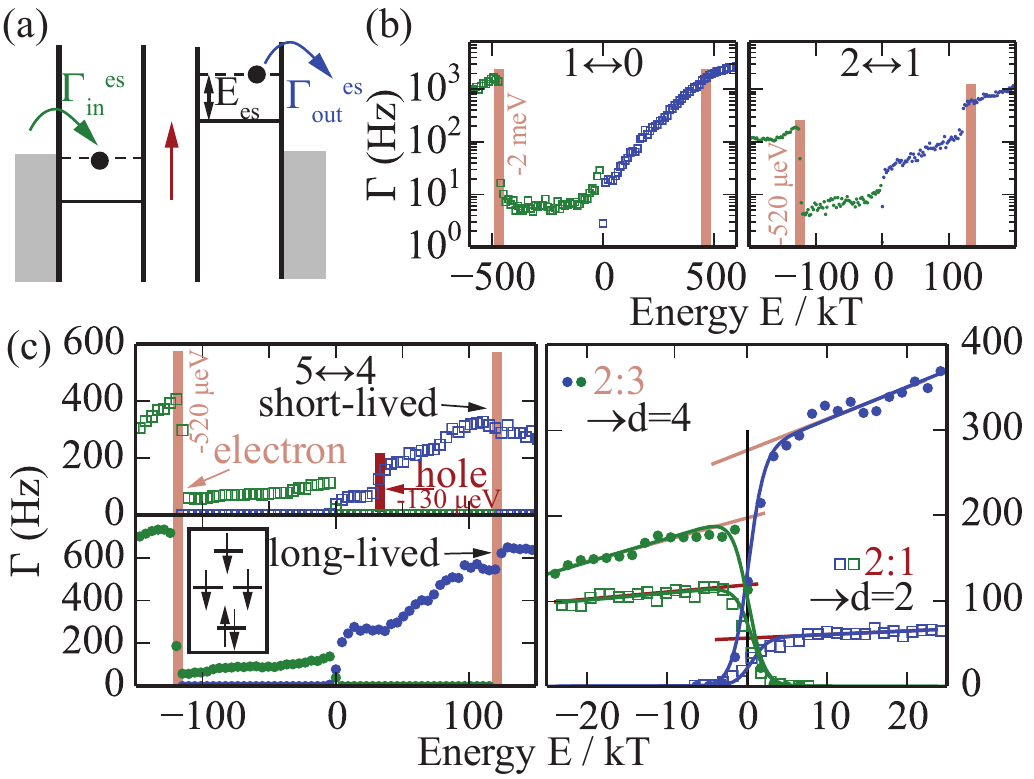}
\caption{Excited state spectroscopy 
($kT\approx3.4~\mu$eV), showing that the feedback technique
allows to infer degeneracy, excitation spectrum and spin-states from one quick
measurement.
The quantum dot states are driven symmetrically as shown in
(a) to measure the tunnelling rates 
$\gin$ and $\gout$.
The data is plotted as open squares (closed circles) in (b) for the 
one-electron (two-electron) state being driven around the Fermi 
energy. $N$-electron excitations are
indicated with red bars and show a short-lived excitation (no step)
for the $N=1$ state and long-lived excitation (step visible) for the $N=2$ state.
The spectroscopy around $\mu_{N=5}$
is shown in (c) for the spin-degenerate (top-left panel, $d=2$) and
the orbital degenerate (bottom-left panel, $d=4$) ground state in open squares and
closed circles, respectively. The right panel shows a zoom
around zero energy at slightly more opened tunnel barrier.}
\label{fig:ExcitedStates}
\end{figure}

The feedback technique also gives access to excitations that are far 
above the ground state energies on the scale of $kT$. 
Excited states of the $N$-electron system can be accessed
\cite{elzerman_single-shot_2004,hanson_single-shot_2005,maclean_energy-dependent_2007,amasha_spin-dependent_2008}, when the dot is initially in the $(N-1)$-electron ground state, 
and the tunnelling-in process brings the system into an 
$N$-electron excited state [see schematic in 
Fig.~\ref{fig:ExcitedStates}(a)]. An example of such a process is shown in 
Fig.~\ref{fig:BField}(a), where a thermally broadened step is seen in 
the tunnelling-in rate into the empty dot (green) close 
to $E=-9kT\approx -\Delta E_Z$, the Zeeman energy, due to the additional 
tunnelling-in channel provided by the spin excitation. A pronounced step is observed
in the tunnelling-out rate at $\Delta E_Z$ in Fig.~\ref{fig:BField}(b) due to the
same spin excitation of the one-electron dot, in which the dot remains after the 
tunnelling-out event.

The analysis of the excited states gives more insight into 
the tunnel coupling of the different states to the lead.
We fit the tunnelling-in rate in Fig.~\ref{fig:BField}(a)
to $[\win f(E) + W_\textrm{0,es} f(E-\ees)][1-\alpha E]$ and 
find $\win = W_\textrm{0,es}=$1:1. Analysing the same excited state in
Fig.~\ref{fig:BField}(b) we find the same ratio for the tunnelling-out rates.
This result agrees with the notion that the orbital single-particle wave function
of the two spin-states are the same.
This precise [relative error $<14\%$ in Fig.~\ref{fig:BField}(a)] 
experimental validation of equal tunnelling rates
is important for charge-to-spin conversion by spin-selective readout
of the charge state 
\cite{elzerman_single-shot_2004,nowack_single-shot_2011,simmons_tunable_2011,amasha_spin-dependent_2008}.

The spin-triplet excitations of the two-electron state provide the possibility to 
compare their tunnel rates
quantitatively. In Fig~\ref{fig:BField}(c), we observe two excitations in the transition between
$(N-1)=1$ and $N=2$,
corresponding to the triplet states $T_0$ and $T_-$, where the tunnel
rate of the $T_-$ state is twice as large as that of the $T_0$
state. The $T_+$ excited state is not observed. These observations
are entirely due to the overlap of the spin-parts of initial and
final states.
The initial state is a statistical mixture of the one-electron quantum dot 
ground state with spin parallel to the magnetic field, and any spin orientation 
in the reservoir, i.e. of $\ket{\downarrow}\otimes\ket{\downarrow}$ and 
$\ket{\downarrow}\otimes\ket{\uparrow}$. Its overlap with the final $T_+$-state 
$\ket{\uparrow\uparrow}$ where both spins in the dot are antiparallel
to the field is therefore zero, and the $T_+$ excited state is not observed
\cite{hanson_determination_2004, hanson_semiconductor_2004, elzerman_excited-state_2004}. 
In contrast, the $T_0$ final state is $(\ket{\uparrow\downarrow} + \ket{\downarrow\uparrow})/\sqrt{2}$
giving a factor of $1/2$ in the squared overlap between initial and final states as compared to
the $T_-$-state $\ket{\downarrow\downarrow}$.

At zero magnetic field, we perform spectroscopy of the orbital states utilizing
the feedback technique.
Fig.~\ref{fig:ExcitedStates}(b) shows an excited state measurement, 
where an excitation of the one-electron (two-electron) dot is filled at $-120kT~(-460kT)\approx-520\mu$eV$~(-2$meV), as seen in the tunnelling-in rate (green).
Interestingly, a corresponding step appears at $+120kT$ in the tunnelling-out rate (blue)
in the two-electron case.
Obviously, the two-electron dot remains 
in the excited state for a longer time than the feedback needs to 
switch the system to higher energy, which is 1.2~ms in our experiment. 
Tunnelling-out from this excited state contributes to the tunnelling-out 
rate at higher energies [see Fig.~\ref{fig:ExcitedStates}(a)]. This 
attests to the long  relaxation time of this lowest excitation, which 
is known to be the spin-triplet state above the spin-singlet ground state 
\cite{hanson_single-shot_2005,amasha_spin-dependent_2008} 
requiring a hyperfine-interaction or spin-orbit interaction mediated 
spin-flip for relaxing to the spin-singlet ground state 
\cite{li_electron_2014,fujisawa_allowed_2002,coish_measurement_2006,khaetskii_spin_2000}. 
A similar mirror step was not seen for the excited state in the left 
panel of Fig.~\ref{fig:ExcitedStates}(b), because orbital excitations 
have lifetimes shorter by many orders of magnitude than those of spin-excitations 
\cite{khaetskii_spin_2000,fujisawa_allowed_2002}.
This example illustrates the remarkable ability
\cite{beveren_spin_2005,fujisawa_allowed_2002} 
of the excited state measurement scheme to distinguish 
spin-excitations from orbital excitations, at zero magnetic field.

We observe a variant of the excited state spectroscopy for the tunnelling 
transition between the four- and the five-electron quantum dot shown in 
the top panel of Fig.~\ref{fig:ExcitedStates}(c). In addition to the fast 
decaying (purely orbital) excitation of the five-electron state visible in 
the tunnelling-in rate (green) at $-120kT\approx520~\mu$eV, a pronounced 
step is observed in the tunnelling-out rate at $+30kT\approx 130\,\mu$eV. 
We interpret this 
step as an excitation of the four-electron $(N-1)$ dot, in which the dot 
remains after the tunnelling-out event, and therefore label it 'hole'. An 
anisotropic two-dimensional harmonic oscillator \cite{fock_bemerkung_1928,darwin_diamagnetism_1931,schuh_algebraic_1985} 
has a single-particle spectrum with two excited state orbitals close in 
energy, but far above the single-particle ground state orbital. It is 
plausible that in our quantum dot, two electrons occupy the lowest orbital 
and that the observed orbital excitation of the four-electron system is a 
single-particle excitation into the higher lying of the nearly-degenerate orbitals.

Next, we turn our attention to the degeneracy measurement of
orbitally degenerate states, which in general have different tunnel couplings
constants.
The orbital excitation of the four-electron system is tuned into 
resonance with the ground state by a suitable change in gate voltages
\cite{amasha_electrical_2008}. The tunnelling rates measured in this situation are shown in the 
lower panel of Fig.~\ref{fig:ExcitedStates}(c),
where the five-electron excitation is still seen in the tunnelling-in rate, but the 
four-electron excitation is resonant with the four-electron ground state,
giving a non-trivial further testing ground for the application of Eq.~\eqref{eq:DetailedBalance}.
The zoom in the right panel of Fig.~\ref{fig:ExcitedStates}(c) shows that
the ratio of tunnelling rates $\win : \wout$ has changed from 2:1 
(open squares, and c.f. Fig.~\ref{fig:EightElectrons}) to 2:3 (filled circles).

The 2:3 ratio of degeneracies of the five- and four-electron quantum dot is
understood within the picture of the single-particle orbital states of the 
anisotropic two-dimensional harmonic oscillator 
\cite{schuh_algebraic_1985}. The single-particle orbital lowest in energy takes two electrons. 
If the next higher orbital energy is twofold degenerate (hence fourfold 
degenerate in total due to the spin-degeneracy of each orbital), the 
three-electron system has a degeneracy of four, the four-electron system 
of six, the five-electron system of four. These numbers are obtained from 
counting the number of ways electrons can be distributed onto the four 
degenerate single-particle states. According to Eq.~\eqref{eq:DetailedBalance}, 
this scenario indeed accounts for the observed ratio 2:3$=$4:6 for the 
tunnelling transition between the four- and the five-electron dot.
Equation~\eqref{eq:DetailedBalance} is valid also for degenerate 
states with different tunnel coupling because it is derived only from 
Detailed Balance and the Second Law of thermodynamics.

The lifetime of the five-electron quantum dot excitation seen at  $-120kT$ in the
top left panel of Fig.\,\ref{fig:ExcitedStates}(d) is short compared to our switching time, if
the four-electron ground state is two-fold spin degenerate. After tuning 
this degeneracy to four, the lifetime of the five-electron excitation has increased as witnessed 
by the mirror step at $+120kT$ in the lower panel
of the same figure. We conclude from the long lifetime that at least one 
relaxation channel requires a spin-flip
[inset of Fig.~\ref{fig:ExcitedStates}(d)], 
which means that the four-electron ground state is a spin-triplet state (Hund's rules).
This demonstrates the connection between ground state degeneracy and 
excitation lifetime and 
agrees with measurements where parallel spin alignment 
of the four-electron ground state has been observed in circular
\cite{ellenberger_excitation_2006, kouwenhoven_excitation_1997,koskinen_hunds_1997}, 
but not in elliptical dots
\cite{sasaki_spin_1998, austing_ellipsoidal_1999}.

In conclusion, we demonstrated ways to precisely measure and tune
the degeneracy of a quantum state, by accessing its tunnelling rates
in a large energy window.
We showed how a quick measurement of the tunnelling rates simultaneously
provides information about degeneracy and spin configuration.
It will be interesting to measure the magnetic field dependence of the 
lifetimes of excited states and in particular, to study the
lifetime of the $T_0$ state in comparison to the lifetime of the $T_-$ state.
Additionally, the feedback loop can be operated in the reverse direction
and thereby realize, for example, a Maxwell's demon setting, which enables cooling 
of the reservoir using information.

\begin{acknowledgments}
We thank Mark Eriksson for stimulating discussions.
We acknowledge the SNF and QSIT for providing the funding which enabled this work.
\end{acknowledgments}
\bibliography{bibliography}
\bibliographystyle{apsrev4-1}
\end{document}